\documentclass[reprint,amsmath,amssymb,aps,twocolumn]{revtex4-2}
\usepackage{float}
\usepackage{appendix}
\usepackage{lipsum}
\usepackage[colorlinks, citecolor=blue,urlcolor=blue,linkcolor=blue,runcolor=filecolor]{hyperref}
\usepackage{url}
\usepackage{amsmath}
\usepackage{subfigure}
\allowdisplaybreaks[4]
\usepackage{graphicx}
\usepackage{dcolumn}
\usepackage{bm}
\begin{document}

\title{Ultralight scalar and  axion dark matter detection with atom interferometers}

\author{Wei Zhao\textsuperscript{1}}
\email{zhaowei164@mails.ucas.ac.cn}
\author{Hui Liu\textsuperscript{1}}
\author{Xitong Mei\textsuperscript{2,3}}
\affiliation{
	\textsuperscript{1}Shandong University of Aeronautics, Binzhou 256600, China\\
	\textsuperscript{2}State Key Laboratory of Magnetic Resonance and Atomic and Molecular Physics,
	Innovation Academy for Precision Measurement Science and Technology, Chinese Academy of Sciences, Wuhan 430071, China	\\ 
   \textsuperscript{3}University of Chinese Academy of Sciences, Beijing 100049, China	}

\date{\today}

\begin{abstract}
The detection of dark matter is a challenging problem in modern physics. The ultralight scalar and axion dark matter could induce the oscillation of the nuclear charge radii and then oscillate the atomic transition frequency by interacting with standard model particles. We compute the differential phase shift caused by the scalar and axion dark matter in a pair of separated atom interferometers and give the proposed constraints on the scalar dark matter coupling parameters $d_g$ and $d_{\hat{m}}$ as well as the axion dark matter coupling parameter $1/f_a$. Our results are expected to improve the current detection level and complement with other experiments. 
\end{abstract}

\maketitle

\section{\label{Introduction}Introduction}
There are many evidences for the existence of dark matter (DM) \cite{Rubin78,Allen03,Clowe2006}. However, no directing DM signal is detected \cite{PandaX-4T2021,LUX-ZEPLIN2023}. Therefore the detection of DM becomes a challenging field in modern physics. The ultralight DM is an important class of DM candidates including the scalar DM, the axion DM and the dark photon DM, where the mass range is $10^{-22} \rm \,eV$ to $1 \rm \,eV$ \cite{Battaglieri17:Report,Safronova18RMP:newphys,WhitePaper2022ScalarVector,Whitepaper2022Axion}. More and more experiments and proposals are put forward to search for these DM candidates, such as accelerometers \cite{Graham16PRD:AccelerometerDM}, atomic clocks \cite{Arvanitaki15PRD,Beloy21Nature:ACDM} and laser interferometers \cite{Vermeulen21:Direct}.

The atom interferometer (AI) is realized by manipulating the atomic matter wave \cite{Cronin09AI}. AIs are widely applied to many fields as a precision measuring tool \cite{Tino21:prospect}, such as the fine structure constant measurement \cite{Morel20FineStructure}, the weak equivalence principle test \cite{Asenbaum20:AI10-12,Elliott23:spaceAI} and the gravitational wave detection \cite{Dimopoulos08PRD:AIGW}. Recently, several AI schemes including a pair of separated AIs are proposed \cite{abend2023terrestrial}, such as AION \cite{Badurina20AION}, MAGIS-100 \cite{Abe2021MAGIS-100}, MIGA \cite{Canuel18MIGA}, ELGAR \cite{Canuel20ELGAR}, ZAIGA \cite{ZAIGA20:Zhan} and AEDGE \cite{El-Neaj20AEDGE}. These developments provide opportunities to detect the ultralight DM with AIs and some proposals are put forward in previous works \cite{Geraci16PRL:AIDMphase,Arvanitaki18PRD:atomicGWDM,Badurina22:RefinedAI,Pumpo22:dilaton,DiPumpo24:AIDM,Du22:AIDM}.

In the ultralight scalar DM model, the scalar DM could interact with the electromagnetic field, the gluon field and the fermion field (electrons and quarks). The corresponding coupling strength described with five coupling parameters $d_{e}$, $d_g$, $d_{m_e}$, $d_{\hat{m}}$ and $d_{\delta m}$.
In the ultralight axion DM model, the axion DM could interact with the gluon field, and the parameter $1/f_a$ describes this interaction strength. 
In this paper, we consider that the oscillation of the nuclear charge radii causes the atom energy level oscillation \cite{Flambaum23:NuclearRadius,Flambaum2023,Dzuba23:NuclearClock,Kim24:QCDaxion}. The atom energy level oscillation could reflect in the differential phase shift of a pair of separated AIs. We compute the induced phase shift by the scalar and axion DM respectively. Due to the oscillation of nuclear charge radii being related with coupling parameters $d_g$ and $d_{\hat{m}}$ of the scalar DM and $1/f_a$ of the axion DM, we could constrain these coupling parameters by a pair of separated AIs. Our results could complement with  some other experiment schemes
\cite{Graham16PRD:AccelerometerDM,Arvanitaki15PRD,Arvanitaki18PRD:atomicGWDM,Beloy21Nature:ACDM,Vermeulen21:Direct,Banerjee2023:NuclearChargeRadii,Flambaum2023,Abel2017:nEDM}. 

The structure of this paper is organized as the following. We introduce the ultralight scalar DM model and axion DM model in Sec. \ref{The dark matter model}.
 In Sec. \ref{The dark matter signal in atom interferometers}, we give the DM signal induced by the scalar DM and the axion DM in a pair of separated AIs. In Sec. \ref{Constraints on the coupling parameters}, we  constrain coupling parameters between DM and standard model particles. In Sec. \ref{Conclusion and discussion}, the conclusion and discussion. 

\section{\label{The dark matter model}The dark matter model}
\subsection{\label {The scalar dark matter model}The scalar dark matter model}
For the scalar DM model, we consider the following interaction Lagrangian density between the scalar DM and standard model particles as \cite{Damour10PRD:DMmodel,Hees18PRD}
\begin{align}
	\mathcal{L}_{\rm int}&=\varphi \Bigg[  \frac{d_{e}}{4 e^2}F_{\mu\nu}F^{\mu\nu}
	-\frac{d_{g}\beta_{3}}{2g_{3}}F^A_{\mu\nu}F^{A\mu\nu}  \notag	\\
	&   - \sum_{i=e,u,d} (d_{m_{i}}+\gamma_{m_{i}}d_{g})m_{i}\psi_{i}\bar{\psi}_{i} \Bigg] \, , \label{3}
\end{align}
where $d_{e}$, $d_{g}$, $d_{m_{e}}$, $d_{m_{u}}$ and $d_{m_{d}}$ are the coupling parameters that scalar DM interact with the electromagnetic field, the gluon field and masses of electrons, up and down quarks. $\beta_{3}$ is the $\beta$-function for $g_{3}$, $m_{i}$ denote  masses of electrons and quarks, $\gamma_{m_{i}}$ is the anomalous dimension due to the renormalization-group running of quark masses, and $\psi_{i}$ are the fermion fields. The $\varphi$ is the scalar DM with the detail form as 
\begin{align}
	\varphi=\varphi_{0} \cos(\omega_\varphi\, t+\delta)\,,
\end{align}
where $\varphi_{0}=\dfrac{\sqrt{2\,\rho_{\rm _{DM}}}}{m_\varphi\,E_{\rm P}}\simeq7 \times10^{-31}\,\rm eV/m_\varphi$ is the scalar DM amplitude,  $\rho_{\rm _{DM}}=0.4\,\rm GeV/cm^3$ is the local DM energy density, $E_{\rm P}$ is the reduced  Planck energy, $\omega_{\varphi}\simeq m_\varphi c^2/\hbar $ is the scalar DM frequency, $m_\varphi$ is the scalar DM mass, $\delta$ is the initial phase.
The Lagrangian could cause the following five physical quantities to oscillate with $\varphi$  as
\begin{align}
	\alpha(\varphi)&=(1+d_{e}\varphi)\alpha 
	\notag \\
	\Lambda_{\rm QCD}(\varphi)&=(1+d_{g}\varphi)\Lambda_{\rm QCD} 
	\notag \\
	m_{i}(\varphi)&=(1+d_{m_{i}}\varphi)m_{i},\quad i=e,u,d \,, \label{scalarchangequantity}
\end{align}
where $\alpha$ is the electromagnetic fine structure constant, and $\Lambda_{\rm QCD}$ is the QCD energy scale.
The mass of quarks could be rewritten as symmetric and antisymmetric combinations
\begin{align}
	\hat{m} =\frac{m_u+m_d}{2}, \,\,\,\, \ 
	\delta m =m_d-m_u \, ,
\end{align}
and the corresponding coupling parameters are 
\begin{align}
	d_{\hat{m}} =\frac{m_ud_{m_u}+m_dd_{m_d}}{m_u+m_d},  \ 
	d_{\delta m} =\frac{m_dd_{m_d}-m_ud_{m_u}}{m_d-m_u} . \label{5}
\end{align}
\subsection {\label{The axion dark matter model}The axion dark matter model}
In this section, we consider the interaction between the axion DM and gluons as  
\begin{align}
\mathcal{L}_{\rm int}=\frac{g^2_{s}}{32\pi^2}\frac{a}{f_a}F^A_{\mu\nu}\tilde{F}^{A\mu\nu}\,,
\end{align}
where $g_{s}$ is a strong coupling constant, $f_a$ is the axion decay
constant, and $\tilde{F}^{A\mu\nu}$ is the dual gluon field strength. The detail form of the aixon field $a$ is 
\begin{align}
a=a_0\cos(\omega_a\,t+\delta)\,,
\end{align}
where $a_0=\sqrt{2\,\rho_{\rm _{DM}}}/m_a\simeq0.0024\,({\rm eV})^2/m_a$ is the axion DM amplitude, $\omega_a \simeq m_a c^2/\hbar$ is the axion DM frequency, $m_a$ is the axion DM mass and $\delta$ is the initial phase. The axion DM could interact with hadrons and further cause changes in the pion mass as \cite{Banerjee2023:NuclearChargeRadii}
\begin{align}
	\frac{\Delta m^2_{\pi}}{m^2_{\pi}}\simeq-\frac{m_um_d}{2(m_u+m_d)^2}\,\theta^2_{\rm eff}(t)=-\frac{m_um_d}{2(m_u+m_d)^2}\frac{1}{f^2_a}a^2\,,\label{mpiAxion}
\end{align}
where $m_u=2.16\,\rm MeV$ and $m_d=4.67\,\rm MeV$ are separately the mass of up and down quarks \cite{ParticleDataGroup}.
\section{\label{The dark matter signal in atom interferometers} The dark matter signal in atom interferometers}
\begin{figure}[htbp]
	\centering
	\includegraphics[width=0.48\textwidth]{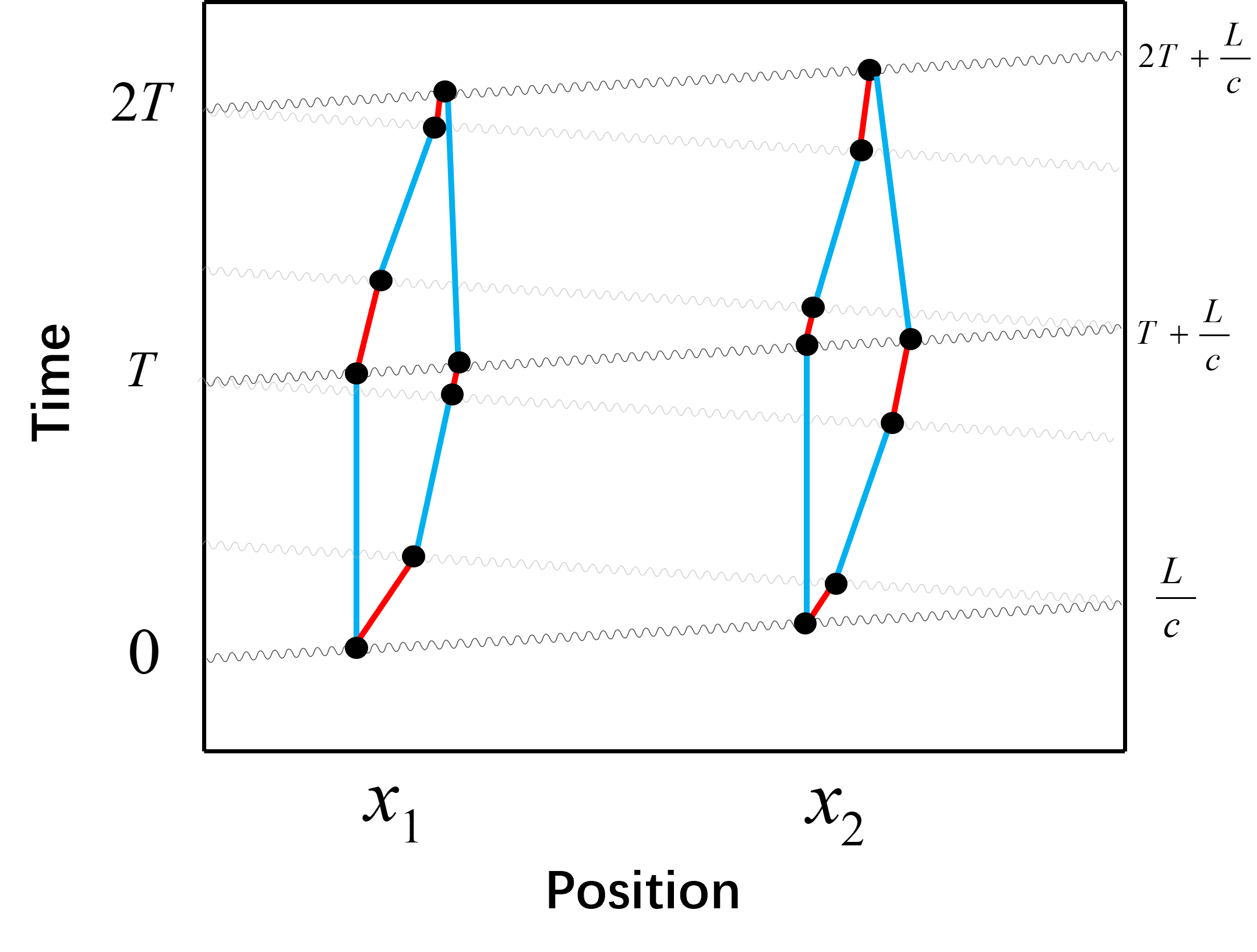}
	\caption{Schematic diagram for a pair of separated AIs. The AIs are located at $x_1$ and $x_2$. The blue and red lines denote atomic paths and the wavy lines denote laser pulses. }
	\label{schematic}
\end{figure}

Here we briefly introduce the principle of a typical $\pi/2-\pi-\pi/2$ light pulse AI \cite{Kasevich91:chuSteven,Kasevich92:chuSteven}. 
First, the atomic packets in the ground state  are divided 
into superposition of ground and excited states with a $\pi/2$-pulse at the time $t=0$. Then a $\pi$-pulse is applied to redirect the ground and excited states at the time $t=T$. Finally, a $\pi/2$-pulse recombines two atomic beams at the time $t=2T$. The phase shift could be obtained by detecting the atom numbers at the ground or excited state. In this paper, we consider a pair of separated  AIs controlled by the common laser with the separated distance being $L\simeq x_2-x_1$ as the Fig. \ref{schematic}. The common laser contributes to cancel out the laser phase noise. The DM signal is obtained by making the differential phase shift measurement between two AIs.

We know the total energy level shift caused by the mass shift which is related with the  nucleus mass and the filed shift which is related with the nuclear charge radius as \cite{krane1991introductory}
\begin{align}
	E_{\rm total}&=E_{\rm MS}+E_{\rm FS}\notag
	\\	&=K\frac{1}{m_N}+F\langle r^2_N\rangle\,,
\end{align}
where $K$ is the mass shift factor, $F$ is the filed shift factor, $m_N$ is the nucleon mass, $\langle r^2_N\rangle$ is the mean squared nuclear charge radius.

The change of the energy level shift for the exited or ground state  is 
\begin{align}
	\Delta E_{\rm total}&=\Delta E_{\rm MS}+\Delta E_{\rm FS}\notag
	\\&=K\frac{1}{m_N}\frac{\Delta m_N}{m_N}+F \langle r^2_N\rangle\frac{\Delta\langle r^2_N\rangle}{\langle r^2_N\rangle}\notag
	\\&\simeq F \langle r^2_N\rangle\frac{\Delta\langle r^2_N\rangle}{\langle r^2_N\rangle}\,.
\end{align}
For the heavy nuclei, such as the Yb atom, the field shift dominates \cite{Banerjee2023:NuclearChargeRadii}.

The corresponding change in the transition frequency is 
\begin{align}
	\Delta \omega_A=\tilde{F} \langle r^2_N\rangle\frac{\Delta\langle r^2_N\rangle}{\langle r^2_N\rangle}\,,
\end{align}
where $\tilde{F}=F_{\rm e}-F_{\rm g}$ is the difference between the field shift factor of the exited state and the ground state. In this paper, we consider the $^1S_0-^3P_0$ clock transition of the Yb atom, where the $\tilde{F}=2\pi\times10.855\,\rm GHz/fm^2$  \cite{Schelfhout21:Ybfieldshift} and $r_N=5.3\,\rm fm $ \cite{Angeli13:NuclearData}. 

The relation of $\langle r^2_N\rangle$ with $f_\pi$ and $m_\pi$ could be found from the Ref. \cite{Banerjee2023:NuclearChargeRadii} as
\begin{align}
	\frac{\Delta\langle r^2_N\rangle}{\langle r^2_N\rangle}
	\simeq \alpha\frac{\Delta f_\pi}{f_\pi}+\beta\frac{\Delta m^2_\pi}{m^2_\pi}
	\simeq \alpha\frac{\Delta \Lambda_{\rm QCD}}{\Lambda_{\rm QCD}}+\beta\frac{\Delta m^2_\pi}{m^2_\pi}\,, \label{rN2}
\end{align}
where $\alpha=-2$\,,  $\beta=-0.2$.

\subsection{\label {The scalar dark matter signal in atom interferometers}The scalar dark matter signal in atom interferometers}
For the scalar dark matter, we could know $m^2_\pi\propto\Lambda_{\rm QCD}\,\hat{m}$ from the Ref. \cite{Ubaldi2010}. So the relation $m^2_\pi$ with
$d_g$ and $d_{\hat{m}}$ is \cite{Banerjee2023:NuclearChargeRadii}
 \begin{align}
 \frac{\Delta m^2_\pi}{m^2_\pi}=\big(d_g+d_{\hat{m}}\big) \,\varphi\,. \label{mpi2}
 \end{align} 
Substituting Eq. (\ref{scalarchangequantity}) and Eq. (\ref{mpi2}) into Eq. (\ref{rN2}), we get
\begin{align}
	\frac{\Delta\langle r^2_N\rangle}{\langle r^2_N\rangle}
	\simeq \big(\alpha+\beta \big)\,d_g\,\varphi+\beta d_{\hat{m}}\,\varphi\,,
\end{align}
and then the change in the transition frequency is 
\begin{align}
	\Delta \omega_A(t)=\tilde{F} \langle r^2_N\rangle \big[\big(\alpha+\beta \big)\,d_g+\beta d_{\hat{m}}\big]\,\varphi_0\cos(\omega_\varphi\,t+\delta)\,.
\end{align}
Due to the interaction between laser pluses and separated AIs, the oscillation in atomic transition frequency will reflect in the differential phase shift of a pair of separated AIs as
\begin{align}
\Phi_{\rm eng}&=\int_{T-(n-1)\frac{L}{c}}^{T+\frac{L}{c}}\Delta\omega_{A}(t)dt-\int_{0}^{\frac{nL}{c}}\Delta\omega_{A}(t)dt
\notag \\
&-\int_{2T-(n-1)\frac{L}{c}}^{2T+\frac{L}{c}}\Delta\omega_{A}(t)dt+\int_{T}^{T+\frac{nL}{c}}\Delta\omega_{A}(t)dt\, .
\end{align}
The differential phase shift  induced by the scalar DM is
\begin{align}
&	\Phi^s_{\rm eng}= 
	\frac{\tilde{F} \langle r^2_N\rangle \big[\big(\alpha+\beta \big)\,d_g+\beta d_{\hat{m}}\big]\,\varphi_0}{\omega_\varphi}
	\notag\\&\times
	\bigg[\sin\big[\omega_\varphi \big(T+\frac{L}{c}\big)+\delta\big]-\sin\big[\omega_\varphi \big(T-(n-1)\frac{L}{c}\big)+\delta\big]
	\notag \\
	&-\sin\big[\frac{\omega_\varphi n L}{c}+\delta\big] +\sin\big[\delta\big]-\sin\big[\omega_\varphi \big(2T+\frac{L}{c}\big)+\delta\big]
	\notag \\
	&+\sin\big[\omega_\varphi \big(2T-(n-1)\frac{L}{c}\big)+\delta\big]
	+\sin\big[\omega_\varphi \big(T+\frac{nL}{c}\big)+\delta\big]
	\notag \\
	&-\sin\big[\omega_\varphi T+\delta\big]\bigg]\, .  \label{transitional phase} 
\end{align}
From the Eq. (\ref{transitional phase}), we could see that the differential phase shift is linearly related with the coupling parameters $d_g$ and $d_{\hat{m}}$. So we could detect the interaction between the scalar DM and the gluon field as well as quarks with AIs. 

\subsection{\label {The axion dark matter signal in atom interferometers}The axion dark matter signal in atom interferometers}
For the axion DM, substituting Eq. (\ref{mpiAxion}) into Eq. (\ref{rN2}), we get
\begin{align}
	\frac{\Delta\langle r^2_N\rangle}{\langle r^2_N\rangle}
	\simeq -\frac{\beta m_um_d}{2(m_u+m_d)^2}\frac{1}{f^2_a}a^2\,,
\end{align}
and the corresponding change in the transition frequency is 
\begin{align}
	\Delta \omega_A=-\tilde{F} \langle r^2_N\rangle \frac{\beta m_um_d}{2(m_u+m_d)^2}\frac{1}{f^2_a}a_0^2\,\cos^2(\omega_a\,t+\delta)\,.
\end{align}
Similarly to the scalar DM,  the axion DM induces the differential phase shift as
\begin{align}
	\Phi^a_{\rm eng}&= 
	-\tilde{F}\,\langle r^2_N\rangle\, \frac{\beta m_um_d}{2(m_u+m_d)^2}\,\frac{a^2_0}{f^2_a}\,\frac{1}{4\,\omega_a}
	\notag \\
	&\times
	\bigg[\sin\big[2\,\omega_a \big(T+\frac{L}{c}\big)+2\,\delta\big]\notag \\
	&-\sin\big[2\,\omega_a \big(T-(n-1)\frac{L}{c}\big)+2\,\delta\big]
	\notag \\
	&-\sin\big[\frac{2\,\omega_a n L}{c}+2\,\delta\big] +\sin\big[2\,\delta\big]\notag \\
	&-\sin\big[2\,\omega_a \big(2T+\frac{L}{c}\big)+2\,\delta\big]
	\notag \\
	&+\sin\big[2\omega_a \big(2T-(n-1)\frac{L}{c}\big)+2\,\delta\big]
	\notag \\
	&+\sin\big[2\,\omega_a\big(T+\frac{nL}{c}\big)+2\delta\big]
	\notag \\
	&-\sin\big[2\,\omega_a T+2\,\delta\big]\bigg]\, . \label{axion transitional phase}
\end{align}

\begin{figure}[thbp]
			\centering
			\includegraphics[width=0.49\textwidth]{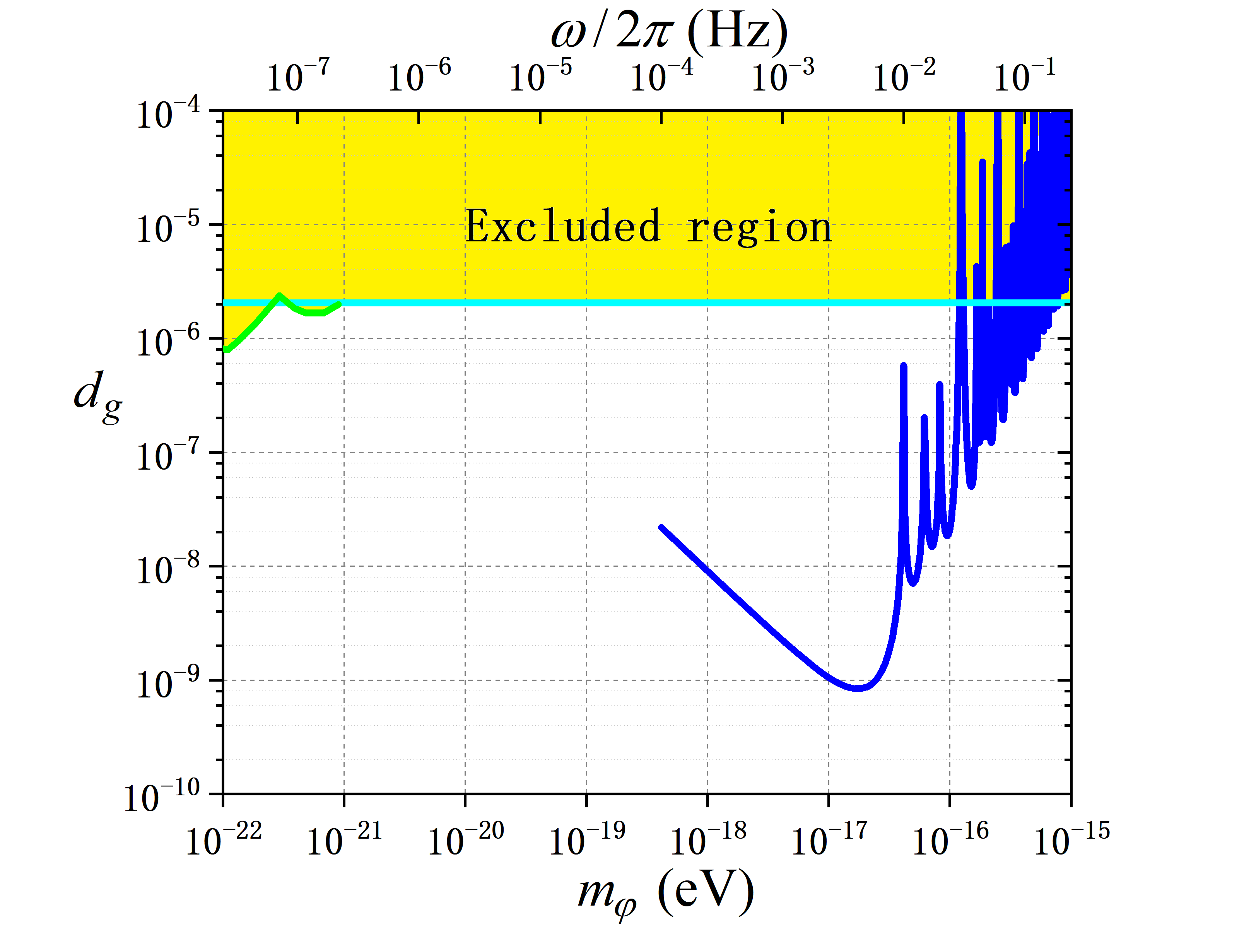}
			\includegraphics[width=0.49\textwidth]{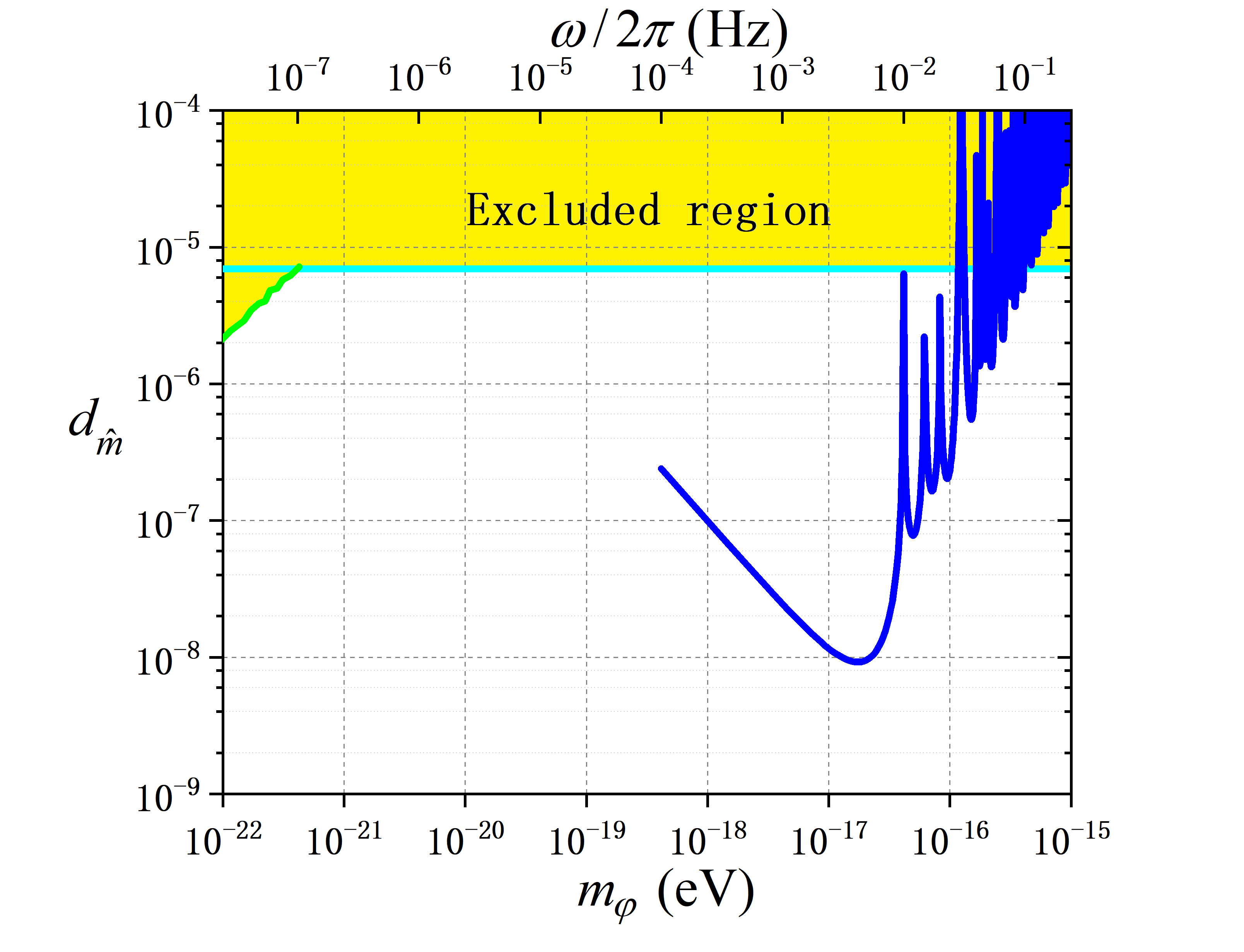}
	\caption{Constraints on the scalar DM coupling parameters $d_g$ and $d_{\hat{m}}$. The blue lines are the constraints on $d_g$ and $d_{\hat{m}}$. The yellow regions are excluded by MICROSCOPE experiments described with cyan lines \cite{Touboul17:MICROSCOPE,Berge18MICROSCOPEConstraints,MICROSCOPE22} and atomic spectroscopy experiments described with green lines \cite{Hees16PRL:AtomicSpectroscopyDM,Kobayashi222,Banerjee2023:NuclearChargeRadii}.}
	\label{constraints on dg and dm}
\end{figure}
\section{\label{Constraints on the coupling parameters}Constraints on the coupling parameters}
According to Refs. \cite{abend2023terrestrial,Badurina20AION,Abe2021MAGIS-100,Canuel18MIGA,Canuel20ELGAR,ZAIGA20:Zhan,El-Neaj20AEDGE}, the technological parameters of AIs are expected to be greatly improved in the future. For a pair of separated AIs in the space, the separated distance is expected to be $L\simeq 10^8\,\rm m$. The phase sensitivity could reach to $10^{-4}\,\rm Hz^{1/2}$. The large momentum transfer number $n\simeq 10^2$ has been demonstrated \cite{Chiow11,Rudolph20}. The free evolution time and integration time could be $T\simeq10^2\,\rm s$ and $t_{int}\simeq10^8\,\rm s$ respectively. For the AI experiments, the shot noise is intrinsic and the gravity gradient noise is important. We ideally only consider this two noise. The shot noise dominates at the high frequency while the  gravity gradient noise dominates below $10^{-4}\rm\,Hz$ \cite{Arvanitaki18PRD:atomicGWDM}.

\subsection{\label{Constraints on the scalar dark matter}Constraints on the scalar dark matter}
We take above technological parameters and only constrain one parameter every time while the other is set to zero. Due to the initial phase $\delta$ being oscillating, we use the  signal amplitude $\bar{\Phi}^s_{\rm eng}=\big[2\int^{2\pi}_0\big(\Phi^s_{\rm eng}\big)^2/2\pi d\delta\big]^{1/2}$ to give the constraints on the coupling parameters. In Fig. \ref{constraints on dg and dm}, we give the constraints on the scalar DM parameters $d_g$ and $d_{\hat{m}}$ with blue lines. The yellow regions are excluded by MICROSCOPE experiments and atomic spectroscopy experiments. We could see that the constraints are better than the result of MICROSCOPE experiments by several orders of magnitude.
Especially at about $m_\varphi=10^{-17}\rm eV$, the constraints are the best. Our results could complement with the result of atomic spectroscopy experiments \cite{Hees16PRL:AtomicSpectroscopyDM,Beloy21Nature:ACDM,Kobayashi222,Banerjee2023:NuclearChargeRadii} and other proposed AI experiments where they give the constraints on coupling parameters $d_e$ and $d_{m_e}$ \cite{Arvanitaki18PRD:atomicGWDM}.

\subsection{\label{Constraints on the axion dark matter}Constraints on the axion dark matter}
\begin{figure}[hbpt]
	\centering
	\includegraphics[width=0.47\textwidth]{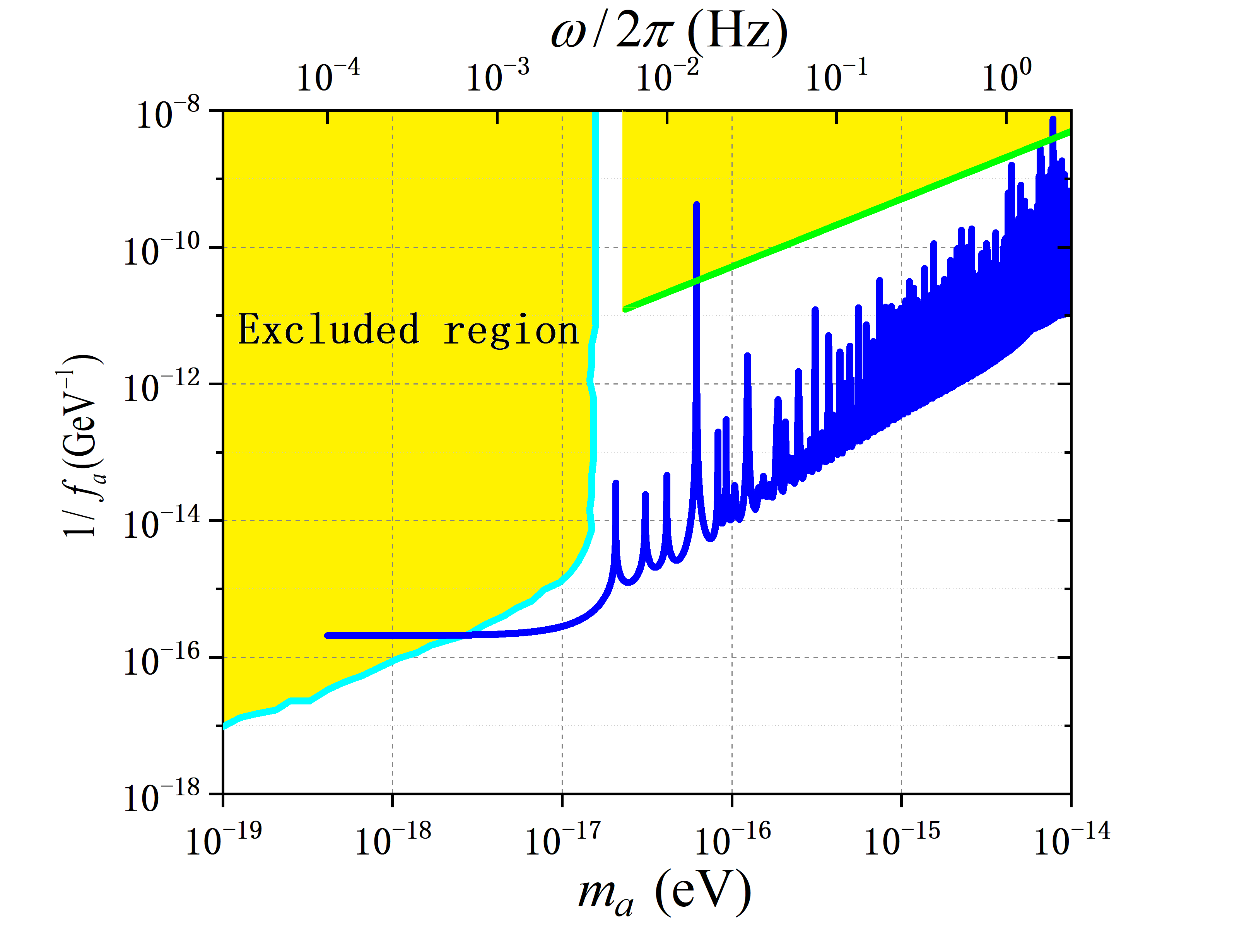}
	\caption{Constraint on the  axion DM coupling parameter $1/f_a$. The blue line is a constraint on the axion DM coupling parameter $1/f_a$. The yellow regions are excluded by the oscillating neutron electric dipole moment experiment described by the cyan line \cite{Abel2017:nEDM} and the atomic clock experiment described by the green line \cite{Flambaum2023}.}
	\label{constraints on fa}
\end{figure}

Similarly to the scalar DM, we also could give constraint on the axion DM coupling parameter $1/f_a$ with $\bar{\Phi}^a_{\rm eng}=\big[2\int^{2\pi}_0\big(\Phi^a_{\rm eng}\big)^2/2\pi d\delta\big]^{1/2}$.  In Fig. \ref{constraints on fa}, the blue line is our proposed constraint. The yellow region is excluded by other experiments \cite{Flambaum2023,Abel2017:nEDM}. This result could complement with the result of the atomic clock experiment \cite{Flambaum2023}  and the oscillating neutron electric dipole moment experiment \cite{Abel2017:nEDM}. 

\section{Conclusion and discussion\label{Conclusion and discussion}}
In this paper, we calculate the differential phase shift caused by the ultralight scalar and axion DM in a pair of separated AIs. The scalar DM signal is linearly related with coupling parameters $d_g$ and $d_{\hat{m}}$ and the axion DM signal is quadratically dependent on the coupling parameter $1/f_a$. The signal strength is also related with the atomic species where the field shift and the nuclear charge radius are different. We further give proposed constraints on the scalar DM coupling parameters $d_g$ and $d_{\hat{m}}$ as well as the axion DM coupling parameters $1/f_a$. The results are expected to improve the current detection level and complement with other experiments.

 \begin{acknowledgements}
  	  This work was supported by the Natural Science Foundation of Shandong Province of China under Grant No. ZR2023QA143, and the Startup Foundation for Doctors of Shandong University of Aeronautics of China under Grant No. 2022Y18.

  \end{acknowledgements}

\bibliography{RefLabelJouranal.bib}

\begin{thebibliography}{54}%
\makeatletter
\providecommand \@ifxundefined [1]{%
 \@ifx{#1\undefined}
}%
\providecommand \@ifnum [1]{%
 \ifnum #1\expandafter \@firstoftwo
 \else \expandafter \@secondoftwo
 \fi
}%
\providecommand \@ifx [1]{%
 \ifx #1\expandafter \@firstoftwo
 \else \expandafter \@secondoftwo
 \fi
}%
\providecommand \natexlab [1]{#1}%
\providecommand \enquote  [1]{``#1''}%
\providecommand \bibnamefont  [1]{#1}%
\providecommand \bibfnamefont [1]{#1}%
\providecommand \citenamefont [1]{#1}%
\providecommand \href@noop [0]{\@secondoftwo}%
\providecommand \href [0]{\begingroup \@sanitize@url \@href}%
\providecommand \@href[1]{\@@startlink{#1}\@@href}%
\providecommand \@@href[1]{\endgroup#1\@@endlink}%
\providecommand \@sanitize@url [0]{\catcode `\\12\catcode `\$12\catcode
  `\&12\catcode `\#12\catcode `\^12\catcode `\_12\catcode `\%12\relax}%
\providecommand \@@startlink[1]{}%
\providecommand \@@endlink[0]{}%
\providecommand \url  [0]{\begingroup\@sanitize@url \@url }%
\providecommand \@url [1]{\endgroup\@href {#1}{\urlprefix }}%
\providecommand \urlprefix  [0]{URL }%
\providecommand \Eprint [0]{\href }%
\providecommand \doibase [0]{https://doi.org/}%
\providecommand \selectlanguage [0]{\@gobble}%
\providecommand \bibinfo  [0]{\@secondoftwo}%
\providecommand \bibfield  [0]{\@secondoftwo}%
\providecommand \translation [1]{[#1]}%
\providecommand \BibitemOpen [0]{}%
\providecommand \bibitemStop [0]{}%
\providecommand \bibitemNoStop [0]{.\EOS\space}%
\providecommand \EOS [0]{\spacefactor3000\relax}%
\providecommand \BibitemShut  [1]{\csname bibitem#1\endcsname}%
\let\auto@bib@innerbib\@empty
\bibitem [{\citenamefont {Rubin}\ \emph {et~al.}(1978)\citenamefont {Rubin},
  \citenamefont {Ford}, \citenamefont {Jr.},\ and\ \citenamefont
  {Thonnard}}]{Rubin78}%
  \BibitemOpen
  \bibfield  {author} {\bibinfo {author} {\bibfnamefont {V.~C.}\ \bibnamefont
  {Rubin}}, \bibinfo {author} {\bibfnamefont {W.~K.}\ \bibnamefont {Ford}},
  \bibinfo {author} {\bibnamefont {Jr.}},\ and\ \bibinfo {author}
  {\bibfnamefont {N.}~\bibnamefont {Thonnard}},\ }\href
  {https://doi.org/10.1086/182804} {\bibfield  {journal} {\bibinfo  {journal}
  {Astrophys. J.}\ }\textbf {\bibinfo {volume} {225}},\ \bibinfo {pages} {L107}
  (\bibinfo {year} {1978})}\BibitemShut {NoStop}%
\bibitem [{\citenamefont {Allen}\ \emph {et~al.}(2003)\citenamefont {Allen},
  \citenamefont {Schmidt}, \citenamefont {Fabian},\ and\ \citenamefont
  {Ebeling}}]{Allen03}%
  \BibitemOpen
  \bibfield  {author} {\bibinfo {author} {\bibfnamefont {S.~W.}\ \bibnamefont
  {Allen}}, \bibinfo {author} {\bibfnamefont {R.~W.}\ \bibnamefont {Schmidt}},
  \bibinfo {author} {\bibfnamefont {A.~C.}\ \bibnamefont {Fabian}},\ and\
  \bibinfo {author} {\bibfnamefont {H.}~\bibnamefont {Ebeling}},\ }\href
  {https://doi.org/10.1046/j.1365-8711.2003.06550.x} {\bibfield  {journal}
  {\bibinfo  {journal} {Mon. Not. R. Astron. Soc.}\ }\textbf {\bibinfo {volume}
  {342}},\ \bibinfo {pages} {287} (\bibinfo {year} {2003})}\BibitemShut
  {NoStop}%
\bibitem [{\citenamefont {Clowe}\ \emph {et~al.}(2006)\citenamefont {Clowe}
  \emph {et~al.}}]{Clowe2006}%
  \BibitemOpen
  \bibfield  {author} {\bibinfo {author} {\bibfnamefont {D.}~\bibnamefont
  {Clowe}} \emph {et~al.},\ }\href {https://doi.org/10.1086/508162} {\bibfield
  {journal} {\bibinfo  {journal} {Astrophys. J.}\ }\textbf {\bibinfo {volume}
  {648}},\ \bibinfo {pages} {L109} (\bibinfo {year} {2006})}\BibitemShut
  {NoStop}%
\bibitem [{\citenamefont {Meng}\ \emph {et~al.}(2021)\citenamefont {Meng} \emph
  {et~al.}}]{PandaX-4T2021}%
  \BibitemOpen
  \bibfield  {author} {\bibinfo {author} {\bibfnamefont {Y.}~\bibnamefont
  {Meng}} \emph {et~al.} (\bibinfo {collaboration} {PandaX-4T Collaboration}),\
  }\href {https://doi.org/10.1103/PhysRevLett.127.261802} {\bibfield  {journal}
  {\bibinfo  {journal} {Phys. Rev. Lett.}\ }\textbf {\bibinfo {volume} {127}},\
  \bibinfo {pages} {261802} (\bibinfo {year} {2021})}\BibitemShut {NoStop}%
\bibitem [{\citenamefont {Aalbers}\ \emph {et~al.}(2023)\citenamefont {Aalbers}
  \emph {et~al.}}]{LUX-ZEPLIN2023}%
  \BibitemOpen
  \bibfield  {author} {\bibinfo {author} {\bibfnamefont {J.}~\bibnamefont
  {Aalbers}} \emph {et~al.} (\bibinfo {collaboration} {LUX-ZEPLIN
  Collaboration}),\ }\href {https://doi.org/10.1103/PhysRevLett.131.041002}
  {\bibfield  {journal} {\bibinfo  {journal} {Phys. Rev. Lett.}\ }\textbf
  {\bibinfo {volume} {131}},\ \bibinfo {pages} {041002} (\bibinfo {year}
  {2023})}\BibitemShut {NoStop}%
\bibitem [{\citenamefont {Battaglieri}\ \emph {et~al.}(2017)\citenamefont
  {Battaglieri} \emph {et~al.}}]{Battaglieri17:Report}%
  \BibitemOpen
  \bibfield  {author} {\bibinfo {author} {\bibfnamefont {M.}~\bibnamefont
  {Battaglieri}} \emph {et~al.},\ }\href@noop {} {\bibfield  {journal}
  {\bibinfo  {journal} {arXiv:1707.04591}\ } (\bibinfo {year}
  {2017})}\BibitemShut {NoStop}%
\bibitem [{\citenamefont {Safronova}\ \emph {et~al.}(2018)\citenamefont
  {Safronova}, \citenamefont {Budker}, \citenamefont {DeMille}, \citenamefont
  {Jackson~Kimball}, \citenamefont {Derevianko},\ and\ \citenamefont
  {Clark}}]{Safronova18RMP:newphys}%
  \BibitemOpen
  \bibfield  {author} {\bibinfo {author} {\bibfnamefont {M.~S.}\ \bibnamefont
  {Safronova}}, \bibinfo {author} {\bibfnamefont {D.}~\bibnamefont {Budker}},
  \bibinfo {author} {\bibfnamefont {D.}~\bibnamefont {DeMille}}, \bibinfo
  {author} {\bibfnamefont {D.~F.}\ \bibnamefont {Jackson~Kimball}}, \bibinfo
  {author} {\bibfnamefont {A.}~\bibnamefont {Derevianko}},\ and\ \bibinfo
  {author} {\bibfnamefont {C.~W.}\ \bibnamefont {Clark}},\ }\href
  {https://doi.org/10.1103/RevModPhys.90.025008} {\bibfield  {journal}
  {\bibinfo  {journal} {Rev. Mod. Phys.}\ }\textbf {\bibinfo {volume} {90}},\
  \bibinfo {pages} {025008} (\bibinfo {year} {2018})}\BibitemShut {NoStop}%
\bibitem [{\citenamefont {Antypas}\ \emph {et~al.}(2022)\citenamefont {Antypas}
  \emph {et~al.}}]{WhitePaper2022ScalarVector}%
  \BibitemOpen
  \bibfield  {author} {\bibinfo {author} {\bibfnamefont {D.}~\bibnamefont
  {Antypas}} \emph {et~al.},\ }\href@noop {} {\bibfield  {journal} {\bibinfo
  {journal} {arXiv:2203.14915}\ } (\bibinfo {year} {2022})}\BibitemShut
  {NoStop}%
\bibitem [{\citenamefont {Adams}\ \emph {et~al.}(2022)\citenamefont {Adams}
  \emph {et~al.}}]{Whitepaper2022Axion}%
  \BibitemOpen
  \bibfield  {author} {\bibinfo {author} {\bibfnamefont {C.~B.}\ \bibnamefont
  {Adams}} \emph {et~al.},\ }\href@noop {} {\bibfield  {journal} {\bibinfo
  {journal} {arXiv:2203.14923}\ } (\bibinfo {year} {2022})}\BibitemShut
  {NoStop}%
\bibitem [{\citenamefont {Graham}\ \emph {et~al.}(2016)\citenamefont {Graham},
  \citenamefont {Kaplan}, \citenamefont {Mardon}, \citenamefont {Rajendran},\
  and\ \citenamefont {Terrano}}]{Graham16PRD:AccelerometerDM}%
  \BibitemOpen
  \bibfield  {author} {\bibinfo {author} {\bibfnamefont {P.~W.}\ \bibnamefont
  {Graham}}, \bibinfo {author} {\bibfnamefont {D.~E.}\ \bibnamefont {Kaplan}},
  \bibinfo {author} {\bibfnamefont {J.}~\bibnamefont {Mardon}}, \bibinfo
  {author} {\bibfnamefont {S.}~\bibnamefont {Rajendran}},\ and\ \bibinfo
  {author} {\bibfnamefont {W.~A.}\ \bibnamefont {Terrano}},\ }\href
  {https://doi.org/10.1103/PhysRevD.93.075029} {\bibfield  {journal} {\bibinfo
  {journal} {Phys. Rev. D}\ }\textbf {\bibinfo {volume} {93}},\ \bibinfo
  {pages} {075029} (\bibinfo {year} {2016})}\BibitemShut {NoStop}%
\bibitem [{\citenamefont {Arvanitaki}\ \emph {et~al.}(2015)\citenamefont
  {Arvanitaki}, \citenamefont {Huang},\ and\ \citenamefont
  {Van~Tilburg}}]{Arvanitaki15PRD}%
  \BibitemOpen
  \bibfield  {author} {\bibinfo {author} {\bibfnamefont {A.}~\bibnamefont
  {Arvanitaki}}, \bibinfo {author} {\bibfnamefont {J.}~\bibnamefont {Huang}},\
  and\ \bibinfo {author} {\bibfnamefont {K.}~\bibnamefont {Van~Tilburg}},\
  }\href {https://doi.org/10.1103/PhysRevD.91.015015} {\bibfield  {journal}
  {\bibinfo  {journal} {Phys. Rev. D}\ }\textbf {\bibinfo {volume} {91}},\
  \bibinfo {pages} {015015} (\bibinfo {year} {2015})}\BibitemShut {NoStop}%
\bibitem [{\citenamefont {Beloy}\ \emph {et~al.}(2021)\citenamefont {Beloy}
  \emph {et~al.}}]{Beloy21Nature:ACDM}%
  \BibitemOpen
  \bibfield  {author} {\bibinfo {author} {\bibfnamefont {K.}~\bibnamefont
  {Beloy}} \emph {et~al.},\ }\href {https://doi.org/10.1038/s41586-021-03253-4}
  {\bibfield  {journal} {\bibinfo  {journal} {Nature}\ }\textbf {\bibinfo
  {volume} {591}},\ \bibinfo {pages} {564} (\bibinfo {year}
  {2021})}\BibitemShut {NoStop}%
\bibitem [{\citenamefont {Vermeulen}\ \emph {et~al.}(2021)\citenamefont
  {Vermeulen} \emph {et~al.}}]{Vermeulen21:Direct}%
  \BibitemOpen
  \bibfield  {author} {\bibinfo {author} {\bibfnamefont {S.~M.}\ \bibnamefont
  {Vermeulen}} \emph {et~al.},\ }\href
  {https://doi.org/10.1038/s41586-021-04031-y} {\bibfield  {journal} {\bibinfo
  {journal} {Nature}\ }\textbf {\bibinfo {volume} {600}},\ \bibinfo {pages}
  {424} (\bibinfo {year} {2021})}\BibitemShut {NoStop}%
\bibitem [{\citenamefont {Cronin}\ \emph {et~al.}(2009)\citenamefont {Cronin},
  \citenamefont {Schmiedmayer},\ and\ \citenamefont {Pritchard}}]{Cronin09AI}%
  \BibitemOpen
  \bibfield  {author} {\bibinfo {author} {\bibfnamefont {A.~D.}\ \bibnamefont
  {Cronin}}, \bibinfo {author} {\bibfnamefont {J.}~\bibnamefont
  {Schmiedmayer}},\ and\ \bibinfo {author} {\bibfnamefont {D.~E.}\ \bibnamefont
  {Pritchard}},\ }\href {https://doi.org/10.1103/RevModPhys.81.1051} {\bibfield
   {journal} {\bibinfo  {journal} {Rev. Mod. Phys.}\ }\textbf {\bibinfo
  {volume} {81}},\ \bibinfo {pages} {1051} (\bibinfo {year}
  {2009})}\BibitemShut {NoStop}%
\bibitem [{\citenamefont {Tino}(2021)}]{Tino21:prospect}%
  \BibitemOpen
  \bibfield  {author} {\bibinfo {author} {\bibfnamefont {G.~M.}\ \bibnamefont
  {Tino}},\ }\href {https://doi.org/10.1088/2058-9565/abd83e} {\bibfield
  {journal} {\bibinfo  {journal} {Quantum Sci. Technol.}\ }\textbf {\bibinfo
  {volume} {6}},\ \bibinfo {pages} {024014} (\bibinfo {year}
  {2021})}\BibitemShut {NoStop}%
\bibitem [{\citenamefont {Morel}\ \emph {et~al.}(2020)\citenamefont {Morel},
  \citenamefont {Yao}, \citenamefont {Clade},\ and\ \citenamefont
  {Guellati-Khelifa}}]{Morel20FineStructure}%
  \BibitemOpen
  \bibfield  {author} {\bibinfo {author} {\bibfnamefont {L.}~\bibnamefont
  {Morel}}, \bibinfo {author} {\bibfnamefont {Z.}~\bibnamefont {Yao}}, \bibinfo
  {author} {\bibfnamefont {P.}~\bibnamefont {Clade}},\ and\ \bibinfo {author}
  {\bibfnamefont {S.}~\bibnamefont {Guellati-Khelifa}},\ }\href
  {https://doi.org/10.1038/s41586-020-2964-7} {\bibfield  {journal} {\bibinfo
  {journal} {Nature}\ }\textbf {\bibinfo {volume} {588}},\ \bibinfo {pages}
  {61} (\bibinfo {year} {2020})}\BibitemShut {NoStop}%
\bibitem [{\citenamefont {Asenbaum}\ \emph {et~al.}(2020)\citenamefont
  {Asenbaum}, \citenamefont {Overstreet}, \citenamefont {Kim}, \citenamefont
  {Curti},\ and\ \citenamefont {Kasevich}}]{Asenbaum20:AI10-12}%
  \BibitemOpen
  \bibfield  {author} {\bibinfo {author} {\bibfnamefont {P.}~\bibnamefont
  {Asenbaum}}, \bibinfo {author} {\bibfnamefont {C.}~\bibnamefont
  {Overstreet}}, \bibinfo {author} {\bibfnamefont {M.}~\bibnamefont {Kim}},
  \bibinfo {author} {\bibfnamefont {J.}~\bibnamefont {Curti}},\ and\ \bibinfo
  {author} {\bibfnamefont {M.~A.}\ \bibnamefont {Kasevich}},\ }\href
  {https://doi.org/10.1103/PhysRevLett.125.191101} {\bibfield  {journal}
  {\bibinfo  {journal} {Phys. Rev. Lett.}\ }\textbf {\bibinfo {volume} {125}},\
  \bibinfo {pages} {191101} (\bibinfo {year} {2020})}\BibitemShut {NoStop}%
\bibitem [{\citenamefont {Elliott}\ \emph {et~al.}(2023)\citenamefont {Elliott}
  \emph {et~al.}}]{Elliott23:spaceAI}%
  \BibitemOpen
  \bibfield  {author} {\bibinfo {author} {\bibfnamefont {E.~R.}\ \bibnamefont
  {Elliott}} \emph {et~al.},\ }\href
  {https://doi.org/10.1038/s41586-023-06645-w} {\bibfield  {journal} {\bibinfo
  {journal} {Nature}\ }\textbf {\bibinfo {volume} {623}},\ \bibinfo {pages}
  {502} (\bibinfo {year} {2023})}\BibitemShut {NoStop}%
\bibitem [{\citenamefont {Dimopoulos}\ \emph {et~al.}(2008)\citenamefont
  {Dimopoulos}, \citenamefont {Graham}, \citenamefont {Hogan}, \citenamefont
  {Kasevich},\ and\ \citenamefont {Rajendran}}]{Dimopoulos08PRD:AIGW}%
  \BibitemOpen
  \bibfield  {author} {\bibinfo {author} {\bibfnamefont {S.}~\bibnamefont
  {Dimopoulos}}, \bibinfo {author} {\bibfnamefont {P.~W.}\ \bibnamefont
  {Graham}}, \bibinfo {author} {\bibfnamefont {J.~M.}\ \bibnamefont {Hogan}},
  \bibinfo {author} {\bibfnamefont {M.~A.}\ \bibnamefont {Kasevich}},\ and\
  \bibinfo {author} {\bibfnamefont {S.}~\bibnamefont {Rajendran}},\ }\href
  {https://doi.org/10.1103/PhysRevD.78.122002} {\bibfield  {journal} {\bibinfo
  {journal} {Phys. Rev. D}\ }\textbf {\bibinfo {volume} {78}},\ \bibinfo
  {pages} {122002} (\bibinfo {year} {2008})}\BibitemShut {NoStop}%
\bibitem [{\citenamefont {Abend}\ \emph {et~al.}(2023)\citenamefont {Abend}
  \emph {et~al.}}]{abend2023terrestrial}%
  \BibitemOpen
  \bibfield  {author} {\bibinfo {author} {\bibfnamefont {S.}~\bibnamefont
  {Abend}} \emph {et~al.},\ }\href@noop {} {\bibfield  {journal} {\bibinfo
  {journal} {arXiv:2310.08183}\ } (\bibinfo {year} {2023})}\BibitemShut
  {NoStop}%
\bibitem [{\citenamefont {Badurina}\ \emph {et~al.}(2020)\citenamefont
  {Badurina} \emph {et~al.}}]{Badurina20AION}%
  \BibitemOpen
  \bibfield  {author} {\bibinfo {author} {\bibfnamefont {L.}~\bibnamefont
  {Badurina}} \emph {et~al.},\ }\href
  {https://doi.org/10.1088/1475-7516/2020/05/011} {\bibfield  {journal}
  {\bibinfo  {journal} {J. Cosmol. Astropart. Phys.}\ }\textbf {\bibinfo
  {volume} {05}}\bibinfo  {number} { (05)},\ \bibinfo {pages}
  {011}}\BibitemShut {NoStop}%
\bibitem [{\citenamefont {Abe}\ \emph {et~al.}(2021)\citenamefont {Abe} \emph
  {et~al.}}]{Abe2021MAGIS-100}%
  \BibitemOpen
\bibfield  {number} {  }\bibfield  {author} {\bibinfo {author} {\bibfnamefont
  {M.}~\bibnamefont {Abe}} \emph {et~al.},\ }\href
  {https://doi.org/10.1088/2058-9565/abf719} {\bibfield  {journal} {\bibinfo
  {journal} {Quantum Sci. Technol.}\ }\textbf {\bibinfo {volume} {6}},\
  \bibinfo {pages} {044003} (\bibinfo {year} {2021})}\BibitemShut {NoStop}%
\bibitem [{\citenamefont {Canuel}\ \emph {et~al.}(2018)\citenamefont {Canuel}
  \emph {et~al.}}]{Canuel18MIGA}%
  \BibitemOpen
  \bibfield  {author} {\bibinfo {author} {\bibfnamefont {B.}~\bibnamefont
  {Canuel}} \emph {et~al.},\ }\href
  {https://doi.org/10.1038/s41598-018-32165-z} {\bibfield  {journal} {\bibinfo
  {journal} {Sci. Rep.}\ }\textbf {\bibinfo {volume} {8}},\ \bibinfo {pages}
  {14064} (\bibinfo {year} {2018})}\BibitemShut {NoStop}%
\bibitem [{\citenamefont {Canuel}\ \emph {et~al.}(2020)\citenamefont {Canuel}
  \emph {et~al.}}]{Canuel20ELGAR}%
  \BibitemOpen
  \bibfield  {author} {\bibinfo {author} {\bibfnamefont {B.}~\bibnamefont
  {Canuel}} \emph {et~al.},\ }\href {https://doi.org/10.1088/1361-6382/aba80e}
  {\bibfield  {journal} {\bibinfo  {journal} {Class. Quantum Grav.}\ }\textbf
  {\bibinfo {volume} {37}},\ \bibinfo {pages} {225017} (\bibinfo {year}
  {2020})}\BibitemShut {NoStop}%
\bibitem [{\citenamefont {Zhan}\ \emph {et~al.}(2020)\citenamefont {Zhan} \emph
  {et~al.}}]{ZAIGA20:Zhan}%
  \BibitemOpen
  \bibfield  {author} {\bibinfo {author} {\bibfnamefont {M.-S.}\ \bibnamefont
  {Zhan}} \emph {et~al.},\ }\href {https://doi.org/10.1142/s0218271819400054}
  {\bibfield  {journal} {\bibinfo  {journal} {Int. J. Mod. Phys. D}\ }\textbf
  {\bibinfo {volume} {29}},\ \bibinfo {pages} {1940005} (\bibinfo {year}
  {2020})}\BibitemShut {NoStop}%
\bibitem [{\citenamefont {El-Neaj}\ \emph {et~al.}(2020)\citenamefont {El-Neaj}
  \emph {et~al.}}]{El-Neaj20AEDGE}%
  \BibitemOpen
  \bibfield  {author} {\bibinfo {author} {\bibfnamefont {Y.~A.}\ \bibnamefont
  {El-Neaj}} \emph {et~al.},\ }\href
  {https://doi.org/10.1140/epjqt/s40507-020-0080-0} {\bibfield  {journal}
  {\bibinfo  {journal} {EPJ Quantum Technol.}\ }\textbf {\bibinfo {volume}
  {7}},\ \bibinfo {pages} {6} (\bibinfo {year} {2020})}\BibitemShut {NoStop}%
\bibitem [{\citenamefont {Geraci}\ and\ \citenamefont
  {Derevianko}(2016)}]{Geraci16PRL:AIDMphase}%
  \BibitemOpen
  \bibfield  {author} {\bibinfo {author} {\bibfnamefont {A.~A.}\ \bibnamefont
  {Geraci}}\ and\ \bibinfo {author} {\bibfnamefont {A.}~\bibnamefont
  {Derevianko}},\ }\href {https://doi.org/10.1103/PhysRevLett.117.261301}
  {\bibfield  {journal} {\bibinfo  {journal} {Phys. Rev. Lett.}\ }\textbf
  {\bibinfo {volume} {117}},\ \bibinfo {pages} {261301} (\bibinfo {year}
  {2016})}\BibitemShut {NoStop}%
\bibitem [{\citenamefont {Arvanitaki}\ \emph {et~al.}(2018)\citenamefont
  {Arvanitaki}, \citenamefont {Graham}, \citenamefont {Hogan}, \citenamefont
  {Rajendran},\ and\ \citenamefont {Van~Tilburg}}]{Arvanitaki18PRD:atomicGWDM}%
  \BibitemOpen
  \bibfield  {author} {\bibinfo {author} {\bibfnamefont {A.}~\bibnamefont
  {Arvanitaki}}, \bibinfo {author} {\bibfnamefont {P.~W.}\ \bibnamefont
  {Graham}}, \bibinfo {author} {\bibfnamefont {J.~M.}\ \bibnamefont {Hogan}},
  \bibinfo {author} {\bibfnamefont {S.}~\bibnamefont {Rajendran}},\ and\
  \bibinfo {author} {\bibfnamefont {K.}~\bibnamefont {Van~Tilburg}},\ }\href
  {https://doi.org/10.1103/PhysRevD.97.075020} {\bibfield  {journal} {\bibinfo
  {journal} {Phys. Rev. D}\ }\textbf {\bibinfo {volume} {97}},\ \bibinfo
  {pages} {075020} (\bibinfo {year} {2018})}\BibitemShut {NoStop}%
\bibitem [{\citenamefont {Badurina}\ \emph {et~al.}(2022)\citenamefont
  {Badurina}, \citenamefont {Blas},\ and\ \citenamefont
  {McCabe}}]{Badurina22:RefinedAI}%
  \BibitemOpen
  \bibfield  {author} {\bibinfo {author} {\bibfnamefont {L.}~\bibnamefont
  {Badurina}}, \bibinfo {author} {\bibfnamefont {D.}~\bibnamefont {Blas}},\
  and\ \bibinfo {author} {\bibfnamefont {C.}~\bibnamefont {McCabe}},\ }\href
  {https://doi.org/10.1103/PhysRevD.105.023006} {\bibfield  {journal} {\bibinfo
   {journal} {Phys. Rev. D}\ }\textbf {\bibinfo {volume} {105}},\ \bibinfo
  {pages} {023006} (\bibinfo {year} {2022})}\BibitemShut {NoStop}%
\bibitem [{\citenamefont {Di~Pumpo}\ \emph {et~al.}(2022)\citenamefont
  {Di~Pumpo}, \citenamefont {Friedrich}, \citenamefont {Geyer}, \citenamefont
  {Ufrecht},\ and\ \citenamefont {Giese}}]{Pumpo22:dilaton}%
  \BibitemOpen
  \bibfield  {author} {\bibinfo {author} {\bibfnamefont {F.}~\bibnamefont
  {Di~Pumpo}}, \bibinfo {author} {\bibfnamefont {A.}~\bibnamefont {Friedrich}},
  \bibinfo {author} {\bibfnamefont {A.}~\bibnamefont {Geyer}}, \bibinfo
  {author} {\bibfnamefont {C.}~\bibnamefont {Ufrecht}},\ and\ \bibinfo {author}
  {\bibfnamefont {E.}~\bibnamefont {Giese}},\ }\href
  {https://doi.org/10.1103/PhysRevD.105.084065} {\bibfield  {journal} {\bibinfo
   {journal} {Phys. Rev. D}\ }\textbf {\bibinfo {volume} {105}},\ \bibinfo
  {pages} {084065} (\bibinfo {year} {2022})}\BibitemShut {NoStop}%
\bibitem [{\citenamefont {Di~Pumpo}\ \emph {et~al.}(2024)\citenamefont
  {Di~Pumpo}, \citenamefont {Friedrich},\ and\ \citenamefont
  {Giese}}]{DiPumpo24:AIDM}%
  \BibitemOpen
  \bibfield  {author} {\bibinfo {author} {\bibfnamefont {F.}~\bibnamefont
  {Di~Pumpo}}, \bibinfo {author} {\bibfnamefont {A.}~\bibnamefont
  {Friedrich}},\ and\ \bibinfo {author} {\bibfnamefont {E.}~\bibnamefont
  {Giese}},\ }\href {https://doi.org/10.1116/5.0175683} {\bibfield  {journal}
  {\bibinfo  {journal} {AVS Quantum Sci.}\ }\textbf {\bibinfo {volume} {6}},\
  \bibinfo {pages} {014404} (\bibinfo {year} {2024})}\BibitemShut {NoStop}%
\bibitem [{\citenamefont {Du}\ \emph {et~al.}(2022)\citenamefont {Du},
  \citenamefont {Murgui}, \citenamefont {Pardo}, \citenamefont {Wang},\ and\
  \citenamefont {Zurek}}]{Du22:AIDM}%
  \BibitemOpen
  \bibfield  {author} {\bibinfo {author} {\bibfnamefont {Y.}~\bibnamefont
  {Du}}, \bibinfo {author} {\bibfnamefont {C.}~\bibnamefont {Murgui}}, \bibinfo
  {author} {\bibfnamefont {K.}~\bibnamefont {Pardo}}, \bibinfo {author}
  {\bibfnamefont {Y.}~\bibnamefont {Wang}},\ and\ \bibinfo {author}
  {\bibfnamefont {K.~M.}\ \bibnamefont {Zurek}},\ }\href
  {https://doi.org/10.1103/PhysRevD.106.095041} {\bibfield  {journal} {\bibinfo
   {journal} {Phys. Rev. D}\ }\textbf {\bibinfo {volume} {106}},\ \bibinfo
  {pages} {095041} (\bibinfo {year} {2022})}\BibitemShut {NoStop}%
\bibitem [{\citenamefont {Flambaum}\ and\ \citenamefont
  {Mansour}(2023)}]{Flambaum23:NuclearRadius}%
  \BibitemOpen
  \bibfield  {author} {\bibinfo {author} {\bibfnamefont {V.~V.}\ \bibnamefont
  {Flambaum}}\ and\ \bibinfo {author} {\bibfnamefont {A.~J.}\ \bibnamefont
  {Mansour}},\ }\href {https://doi.org/10.1103/PhysRevLett.131.113004}
  {\bibfield  {journal} {\bibinfo  {journal} {Phys. Rev. Lett.}\ }\textbf
  {\bibinfo {volume} {131}},\ \bibinfo {pages} {113004} (\bibinfo {year}
  {2023})}\BibitemShut {NoStop}%
\bibitem [{\citenamefont {Flambaum}\ and\ \citenamefont
  {Samsonov}(2023)}]{Flambaum2023}%
  \BibitemOpen
  \bibfield  {author} {\bibinfo {author} {\bibfnamefont {V.~V.}\ \bibnamefont
  {Flambaum}}\ and\ \bibinfo {author} {\bibfnamefont {I.~B.}\ \bibnamefont
  {Samsonov}},\ }\href {https://doi.org/10.1103/PhysRevD.108.075022} {\bibfield
   {journal} {\bibinfo  {journal} {Phys. Rev. D}\ }\textbf {\bibinfo {volume}
  {108}},\ \bibinfo {pages} {075022} (\bibinfo {year} {2023})}\BibitemShut
  {NoStop}%
\bibitem [{\citenamefont {Dzuba}\ and\ \citenamefont
  {Flambaum}(2023)}]{Dzuba23:NuclearClock}%
  \BibitemOpen
  \bibfield  {author} {\bibinfo {author} {\bibfnamefont {V.~A.}\ \bibnamefont
  {Dzuba}}\ and\ \bibinfo {author} {\bibfnamefont {V.~V.}\ \bibnamefont
  {Flambaum}},\ }\href {https://doi.org/10.1103/PhysRevLett.131.263002}
  {\bibfield  {journal} {\bibinfo  {journal} {Phys. Rev. Lett.}\ }\textbf
  {\bibinfo {volume} {131}},\ \bibinfo {pages} {263002} (\bibinfo {year}
  {2023})}\BibitemShut {NoStop}%
\bibitem [{\citenamefont {Kim}\ and\ \citenamefont
  {Perez}(2024)}]{Kim24:QCDaxion}%
  \BibitemOpen
  \bibfield  {author} {\bibinfo {author} {\bibfnamefont {H.}~\bibnamefont
  {Kim}}\ and\ \bibinfo {author} {\bibfnamefont {G.}~\bibnamefont {Perez}},\
  }\href {https://doi.org/10.1103/PhysRevD.109.015005} {\bibfield  {journal}
  {\bibinfo  {journal} {Phys. Rev. D}\ }\textbf {\bibinfo {volume} {109}},\
  \bibinfo {pages} {015005} (\bibinfo {year} {2024})}\BibitemShut {NoStop}%
\bibitem [{\citenamefont {Banerjee}\ \emph {et~al.}(2023)\citenamefont
  {Banerjee} \emph {et~al.}}]{Banerjee2023:NuclearChargeRadii}%
  \BibitemOpen
  \bibfield  {author} {\bibinfo {author} {\bibfnamefont {A.}~\bibnamefont
  {Banerjee}} \emph {et~al.},\ }\href@noop {} {\bibfield  {journal} {\bibinfo
  {journal} {arXiv:2301.10784}\ } (\bibinfo {year} {2023})}\BibitemShut
  {NoStop}%
\bibitem [{\citenamefont {Abel}\ \emph {et~al.}(2017)\citenamefont {Abel} \emph
  {et~al.}}]{Abel2017:nEDM}%
  \BibitemOpen
  \bibfield  {author} {\bibinfo {author} {\bibfnamefont {C.}~\bibnamefont
  {Abel}} \emph {et~al.},\ }\href {https://doi.org/10.1103/PhysRevX.7.041034}
  {\bibfield  {journal} {\bibinfo  {journal} {Phys. Rev. X}\ }\textbf {\bibinfo
  {volume} {7}},\ \bibinfo {pages} {041034} (\bibinfo {year}
  {2017})}\BibitemShut {NoStop}%
\bibitem [{\citenamefont {Damour}\ and\ \citenamefont
  {Donoghue}(2010)}]{Damour10PRD:DMmodel}%
  \BibitemOpen
  \bibfield  {author} {\bibinfo {author} {\bibfnamefont {T.}~\bibnamefont
  {Damour}}\ and\ \bibinfo {author} {\bibfnamefont {J.~F.}\ \bibnamefont
  {Donoghue}},\ }\href {https://doi.org/10.1103/PhysRevD.82.084033} {\bibfield
  {journal} {\bibinfo  {journal} {Phys. Rev. D}\ }\textbf {\bibinfo {volume}
  {82}},\ \bibinfo {pages} {084033} (\bibinfo {year} {2010})}\BibitemShut
  {NoStop}%
\bibitem [{\citenamefont {Hees}\ \emph {et~al.}(2018)\citenamefont {Hees},
  \citenamefont {Minazzoli}, \citenamefont {Savalle}, \citenamefont {Stadnik},\
  and\ \citenamefont {Wolf}}]{Hees18PRD}%
  \BibitemOpen
  \bibfield  {author} {\bibinfo {author} {\bibfnamefont {A.}~\bibnamefont
  {Hees}}, \bibinfo {author} {\bibfnamefont {O.}~\bibnamefont {Minazzoli}},
  \bibinfo {author} {\bibfnamefont {E.}~\bibnamefont {Savalle}}, \bibinfo
  {author} {\bibfnamefont {Y.~V.}\ \bibnamefont {Stadnik}},\ and\ \bibinfo
  {author} {\bibfnamefont {P.}~\bibnamefont {Wolf}},\ }\href
  {https://doi.org/10.1103/PhysRevD.98.064051} {\bibfield  {journal} {\bibinfo
  {journal} {Phys. Rev. D}\ }\textbf {\bibinfo {volume} {98}},\ \bibinfo
  {pages} {064051} (\bibinfo {year} {2018})}\BibitemShut {NoStop}%
\bibitem [{\citenamefont {Workman}\ \emph {et~al.}(2022)\citenamefont {Workman}
  \emph {et~al.}}]{ParticleDataGroup}%
  \BibitemOpen
  \bibfield  {author} {\bibinfo {author} {\bibfnamefont {R.~L.}\ \bibnamefont
  {Workman}} \emph {et~al.} (\bibinfo {collaboration} {Particle Data Group}),\
  }\href {https://doi.org/10.1093/ptep/ptac097} {\bibfield  {journal} {\bibinfo
   {journal} {Prog. Theor. Exp. Phys.}\ }\textbf {\bibinfo {volume} {2022}},\
  \bibinfo {pages} {083C01} (\bibinfo {year} {2022})}\BibitemShut {NoStop}%
\bibitem [{\citenamefont {Kasevich}\ and\ \citenamefont
  {Chu}(1991)}]{Kasevich91:chuSteven}%
  \BibitemOpen
  \bibfield  {author} {\bibinfo {author} {\bibfnamefont {M.}~\bibnamefont
  {Kasevich}}\ and\ \bibinfo {author} {\bibfnamefont {S.}~\bibnamefont {Chu}},\
  }\href {https://doi.org/10.1103/PhysRevLett.67.181} {\bibfield  {journal}
  {\bibinfo  {journal} {Phys. Rev. Lett.}\ }\textbf {\bibinfo {volume} {67}},\
  \bibinfo {pages} {181} (\bibinfo {year} {1991})}\BibitemShut {NoStop}%
\bibitem [{\citenamefont {Kasevich}\ and\ \citenamefont
  {Chu}(1992)}]{Kasevich92:chuSteven}%
  \BibitemOpen
  \bibfield  {author} {\bibinfo {author} {\bibfnamefont {M.}~\bibnamefont
  {Kasevich}}\ and\ \bibinfo {author} {\bibfnamefont {S.}~\bibnamefont {Chu}},\
  }\href {https://doi.org/10.1007/bf00325375} {\bibfield  {journal} {\bibinfo
  {journal} {Appl. Phys. B}\ }\textbf {\bibinfo {volume} {54}},\ \bibinfo
  {pages} {321} (\bibinfo {year} {1992})}\BibitemShut {NoStop}%
\bibitem [{\citenamefont {Krane}(1991)}]{krane1991introductory}%
  \BibitemOpen
  \bibfield  {author} {\bibinfo {author} {\bibfnamefont {K.}~\bibnamefont
  {Krane}},\ }\href {https://books.google.de/books?id=BNtVEAAAQBAJ} {\emph
  {\bibinfo {title} {Introductory Nuclear Physics}}}\ (\bibinfo  {publisher}
  {Wiley},\ \bibinfo {year} {1991})\BibitemShut {NoStop}%
\bibitem [{\citenamefont {Schelfhout}\ and\ \citenamefont
  {McFerran}(2021)}]{Schelfhout21:Ybfieldshift}%
  \BibitemOpen
  \bibfield  {author} {\bibinfo {author} {\bibfnamefont {J.~S.}\ \bibnamefont
  {Schelfhout}}\ and\ \bibinfo {author} {\bibfnamefont {J.~J.}\ \bibnamefont
  {McFerran}},\ }\href {https://doi.org/10.1103/PhysRevA.104.022806} {\bibfield
   {journal} {\bibinfo  {journal} {Phys. Rev. A}\ }\textbf {\bibinfo {volume}
  {104}},\ \bibinfo {pages} {022806} (\bibinfo {year} {2021})}\BibitemShut
  {NoStop}%
\bibitem [{\citenamefont {Angeli}\ and\ \citenamefont
  {Marinova}(2013)}]{Angeli13:NuclearData}%
  \BibitemOpen
  \bibfield  {author} {\bibinfo {author} {\bibfnamefont {I.}~\bibnamefont
  {Angeli}}\ and\ \bibinfo {author} {\bibfnamefont {K.~P.}\ \bibnamefont
  {Marinova}},\ }\href
  {https://doi.org/https://doi.org/10.1016/j.adt.2011.12.006} {\bibfield
  {journal} {\bibinfo  {journal} {At. Data Nucl. Data Tables}\ }\textbf
  {\bibinfo {volume} {99}},\ \bibinfo {pages} {69} (\bibinfo {year}
  {2013})}\BibitemShut {NoStop}%
\bibitem [{\citenamefont {Ubaldi}(2010)}]{Ubaldi2010}%
  \BibitemOpen
  \bibfield  {author} {\bibinfo {author} {\bibfnamefont {L.}~\bibnamefont
  {Ubaldi}},\ }\href {https://doi.org/10.1103/PhysRevD.81.025011} {\bibfield
  {journal} {\bibinfo  {journal} {Phys. Rev. D}\ }\textbf {\bibinfo {volume}
  {81}},\ \bibinfo {pages} {025011} (\bibinfo {year} {2010})}\BibitemShut
  {NoStop}%
\bibitem [{\citenamefont {Touboul}\ \emph {et~al.}(2017)\citenamefont {Touboul}
  \emph {et~al.}}]{Touboul17:MICROSCOPE}%
  \BibitemOpen
  \bibfield  {author} {\bibinfo {author} {\bibfnamefont {P.}~\bibnamefont
  {Touboul}} \emph {et~al.} (\bibinfo {collaboration} {MICROSCOPE
  Collaboration}),\ }\href {https://doi.org/10.1103/PhysRevLett.119.231101}
  {\bibfield  {journal} {\bibinfo  {journal} {Phys. Rev. Lett.}\ }\textbf
  {\bibinfo {volume} {119}},\ \bibinfo {pages} {231101} (\bibinfo {year}
  {2017})}\BibitemShut {NoStop}%
\bibitem [{\citenamefont {Berg\'e}\ \emph {et~al.}(2018)\citenamefont
  {Berg\'e}, \citenamefont {Brax}, \citenamefont {M\'etris}, \citenamefont
  {Pernot-Borr\`as}, \citenamefont {Touboul},\ and\ \citenamefont
  {Uzan}}]{Berge18MICROSCOPEConstraints}%
  \BibitemOpen
  \bibfield  {author} {\bibinfo {author} {\bibfnamefont {J.}~\bibnamefont
  {Berg\'e}}, \bibinfo {author} {\bibfnamefont {P.}~\bibnamefont {Brax}},
  \bibinfo {author} {\bibfnamefont {G.}~\bibnamefont {M\'etris}}, \bibinfo
  {author} {\bibfnamefont {M.}~\bibnamefont {Pernot-Borr\`as}}, \bibinfo
  {author} {\bibfnamefont {P.}~\bibnamefont {Touboul}},\ and\ \bibinfo {author}
  {\bibfnamefont {J.-P.}\ \bibnamefont {Uzan}},\ }\href
  {https://doi.org/10.1103/PhysRevLett.120.141101} {\bibfield  {journal}
  {\bibinfo  {journal} {Phys. Rev. Lett.}\ }\textbf {\bibinfo {volume} {120}},\
  \bibinfo {pages} {141101} (\bibinfo {year} {2018})}\BibitemShut {NoStop}%
\bibitem [{\citenamefont {Touboul}\ \emph {et~al.}(2022)\citenamefont {Touboul}
  \emph {et~al.}}]{MICROSCOPE22}%
  \BibitemOpen
  \bibfield  {author} {\bibinfo {author} {\bibfnamefont {P.}~\bibnamefont
  {Touboul}} \emph {et~al.} (\bibinfo {collaboration} {MICROSCOPE
  Collaboration}),\ }\href {https://doi.org/10.1103/PhysRevLett.129.121102}
  {\bibfield  {journal} {\bibinfo  {journal} {Phys. Rev. Lett.}\ }\textbf
  {\bibinfo {volume} {129}},\ \bibinfo {pages} {121102} (\bibinfo {year}
  {2022})}\BibitemShut {NoStop}%
\bibitem [{\citenamefont {Hees}\ \emph {et~al.}(2016)\citenamefont {Hees},
  \citenamefont {Guena}, \citenamefont {Abgrall}, \citenamefont {Bize},\ and\
  \citenamefont {Wolf}}]{Hees16PRL:AtomicSpectroscopyDM}%
  \BibitemOpen
  \bibfield  {author} {\bibinfo {author} {\bibfnamefont {A.}~\bibnamefont
  {Hees}}, \bibinfo {author} {\bibfnamefont {J.}~\bibnamefont {Guena}},
  \bibinfo {author} {\bibfnamefont {M.}~\bibnamefont {Abgrall}}, \bibinfo
  {author} {\bibfnamefont {S.}~\bibnamefont {Bize}},\ and\ \bibinfo {author}
  {\bibfnamefont {P.}~\bibnamefont {Wolf}},\ }\href
  {https://doi.org/10.1103/PhysRevLett.117.061301} {\bibfield  {journal}
  {\bibinfo  {journal} {Phys. Rev. Lett.}\ }\textbf {\bibinfo {volume} {117}},\
  \bibinfo {pages} {061301} (\bibinfo {year} {2016})}\BibitemShut {NoStop}%
\bibitem [{\citenamefont {Kobayashi}\ \emph {et~al.}(2022)\citenamefont
  {Kobayashi} \emph {et~al.}}]{Kobayashi222}%
  \BibitemOpen
  \bibfield  {author} {\bibinfo {author} {\bibfnamefont {T.}~\bibnamefont
  {Kobayashi}} \emph {et~al.},\ }\href
  {https://doi.org/10.1103/PhysRevLett.129.241301} {\bibfield  {journal}
  {\bibinfo  {journal} {Phys. Rev. Lett.}\ }\textbf {\bibinfo {volume} {129}},\
  \bibinfo {pages} {241301} (\bibinfo {year} {2022})}\BibitemShut {NoStop}%
\bibitem [{\citenamefont {Chiow}\ \emph {et~al.}(2011)\citenamefont {Chiow},
  \citenamefont {Kovachy}, \citenamefont {Chien},\ and\ \citenamefont
  {Kasevich}}]{Chiow11}%
  \BibitemOpen
  \bibfield  {author} {\bibinfo {author} {\bibfnamefont {S.~W.}\ \bibnamefont
  {Chiow}}, \bibinfo {author} {\bibfnamefont {T.}~\bibnamefont {Kovachy}},
  \bibinfo {author} {\bibfnamefont {H.~C.}\ \bibnamefont {Chien}},\ and\
  \bibinfo {author} {\bibfnamefont {M.~A.}\ \bibnamefont {Kasevich}},\ }\href
  {https://doi.org/10.1103/PhysRevLett.107.130403} {\bibfield  {journal}
  {\bibinfo  {journal} {Phys. Rev. Lett.}\ }\textbf {\bibinfo {volume} {107}},\
  \bibinfo {pages} {130403} (\bibinfo {year} {2011})}\BibitemShut {NoStop}%
\bibitem [{\citenamefont {Rudolph}\ \emph {et~al.}(2020)\citenamefont {Rudolph}
  \emph {et~al.}}]{Rudolph20}%
  \BibitemOpen
  \bibfield  {author} {\bibinfo {author} {\bibfnamefont {J.}~\bibnamefont
  {Rudolph}} \emph {et~al.},\ }\href
  {https://doi.org/10.1103/PhysRevLett.124.083604} {\bibfield  {journal}
  {\bibinfo  {journal} {Phys. Rev. Lett.}\ }\textbf {\bibinfo {volume} {124}},\
  \bibinfo {pages} {083604} (\bibinfo {year} {2020})}\BibitemShut {NoStop}%
\end{thebibliography}%

\end{document}